\documentclass[runningheads]{llncs}
\usepackage[T1]{fontenc}

\usepackage{hyperref}
\usepackage{url}
\usepackage{graphicx}
\usepackage{amsmath}
\usepackage{amsfonts}
\usepackage{array}
\usepackage{tikz}
\usepackage{tcolorbox}
\newcolumntype{P}[1]{>{\centering\arraybackslash}p{#1}}

\begin{document}
\title{Chan-Vese Attention U-Net: An attention mechanism for robust segmentation.}
\titlerunning{Chan-Vese Attention U-Net}

\author{Nicolas Makaroff \inst{1} \and Laurent D. Cohen \inst{1}}
\authorrunning{Makaroff et al.}

\institute{CEREMADE, UMR CNRS 7534, University Paris Dauphine, PSL Research University, 75775 Paris, France
\email{makaroff@ceremade.dauphine.fr}\\
}
\maketitle              
\begin{abstract}
When studying the results of a segmentation algorithm using convolutional neural networks, one wonders about the reliability and consistency of the results. This leads to questioning the possibility of using such an algorithm in applications where there is little room for doubt. We propose in this paper a new attention gate based on the use of Chan-Vese energy minimization to control more precisely the segmentation masks given by a standard CNN architecture such as the U-Net model. This mechanism allows to obtain a constraint on the segmentation based on the resolution of a PDE. The study of the results allows us to observe the spatial information retained by the neural network on the region of interest and obtains competitive results on the binary segmentation. We illustrate the efficiency of this approach for medical image segmentation on a database of MRI brain images.

\end{abstract}
\section{Introduction}

Medical image segmentation is a crucial task that requires significant time and effort from medical experts. Although various solutions, including convolutional neural networks (CNNs), have been proposed to automate this process, the need for efficient and reliable methods still exists. 

While CNNs have shown promising results in medical image segmentation, their lack of transparency and the sensitive nature of medical data raise concerns regarding their applicability in real-world hospital settings, especially for medical staff who are not trained in machine learning.

Convolutional neural networks have revolutionised the study of images for both classification and segmentation, the latter being the objective of interest here. Two architectures stand out among all those studied over the years, the fully convolutional neural network \cite{conf/cvpr/LongSD15} and the famous U-Net \cite{ronneberger2015u}. These architectures have been tested many times on various applications such as MRI segmentation of the brain \cite{kleesiek2016deep} or heart \cite{2018}, CT scans of organs in the thoracic cavity \cite{8759212}. Numerous modifications have also been made to improve the efficiency of these complex structures and in the medical field especially the U-Net has proven to be very versatile for many segmentation tasks.
The rest of the paper is organised as follows. In Section \ref{sec:method} we introduce our experimental method for Fast Marching Energy CNN. In Section \ref{sec:experiments} we present the main results of our experiments and provide a discussion around our work.

\subsection{Related work}

Several attempts have been made to integrate geometric or topological properties in the neural network to incorporate information beyond adjacent pixels for segmentation tasks. In 2019, \cite{hatamizadeh2019deep} proposed a lesion segmentation method based on active contours using a U-Net-like neural network. The network predicts a segmentation mask, which is refined using a generalisation of the active contour problem from \cite{Chan2001ActiveCW}. Similarly, In 2020, \cite{zhang2020deep} presented a model, where the neural network predicts the parameters for initialising the active contour model and an initial contour. Learning is achieved by combining the error produced by the neural network and that produced during active contour usage. In 2021, \cite{ma2020learning} proposed a fully integrated geodesic active contour model, where the neural network learns to minimise the energy functional of the model. In this encoder-decoder network, the output is a contour map instead of a probability map for segmentation, based on the active contour method proposed by \cite{Caselles95geodesicactive}. Although these methods have shown promise, there is still room for improvement in incorporating geometric and topological properties more effectively and seamlessly into deep learning-based segmentation approaches.

\subsection{Contributions}

The main contribution of this paper is the development of a novel hybrid segmentation method that combines deep learning with classical functional energy minimization techniques, specifically designed for medical image segmentation applications. Our method features a new attention gate, the $\textit{Chan-Vese Attention Gate}$, which integrates information from the level sets method of the well-established Chan-Vese functional \cite{Chan2001ActiveCW}. Unlike traditional deep learning methods that rely solely on the neural network to improve image segmentation, our approach leverages resolution information to achieve more accurate results.

To demonstrate the effectiveness of our method, we conducted comprehensive experiments on the TCGA\_LGG database \cite{tcga_lgg}, a repository of brain images for the study of lower grade gliomas. Given the sensitive nature of medical image segmentation, it is crucial to ensure the validity of our results. Our approach achieved at least equivalent results to previous networks while remaining simple to optimise. The training time for our method is only 5\% longer than the traditional approach, which is a minor increase considering the benefits. Furthermore, the difference in computation time is imperceptible during inference. Overall, our approach represents a significant advancement in medical image segmentation, offering a more accurate and efficacious solution for this critical field.

\section{Methodology}
\label{sec:method}

\paragraph{The U-Net architecture}

The U-Net architecture is widely used for medical image segmentation. It maintains the structure of an image during transformation from an image to a vector and back to an image using features extracted during the contraction phase. The architecture takes the shape of a "U" with three parts: contraction, transition, and expansion. The contraction applies several blocks with convolution and pooling layers, doubling the feature maps at each stage. The transition uses convolution layers, while the expansion uses convolution and up-sampling layers. The information is recovered during the contraction to reconstruct the image, with the same number of expansion blocks as contraction blocks. The final output is obtained through the final convolutional layer.

\paragraph{Attention Gate in U-Net architecture}

The authors of \cite{1804.03999} presented in their paper a new attention gate especially for the case of the CNN and in particular for the U-Net. The architecture of the U-Net remains the same except for the expansion part. In this part, an attention mechanism is integrated between each block. For each block, the input and the information coming from the connections of the corresponding contraction part block pass through an attention block. The input is up-sampled in parallel and finally the two results are concatenated and sent to the convolution block.

\paragraph{Chan-Vese Energy Minimization}

Presented in \cite{Chan2001ActiveCW}, Chan-Vese's method is used to segment a binary image. Let $I$ be the given grayscale image on a domain $\Omega$ to be segmented. The Chan Vese method looks for a piece-wise constant approximation of an image where there are 2 regions separated by an unknown boundary curve C. This is obtained through the minimization of the following energy depending on curve C and the constant values c1 and c2 inside and outside the curve:

\begin{multline}
    E(C,c_1,c_2) = \mu\times\text{Length}(C)+\nu\times\text{Area}(inside(C)) \\
    + \lambda_1 \int_{inside(C)}|I(x,y)-c_1|^2 \mathrm{d}x\mathrm{d}y + \lambda_2 \int_{outside(C)}|I(x,y)-c_2|^2 \mathrm{d}x\mathrm{d}y.
\end{multline}
\noindent
Energy minimization is simplified by replacing the curve $C$ with a level set function $\phi$. The inside region is then the set where 
$\phi>0$ and the outside region the set where $\phi<0$. With the help of the Heavyside function $H$, the energy becomes:

\begin{multline}
    F(c_1,c_2, \phi) =  \mu\int_{\Omega}\delta(\phi(x))|\nabla\phi(x)|dx + \nu\int_\Omega H(\phi(x))dx \\
    + \lambda_1\int_\Omega |I(x) - c_1|^2H(\phi(x))dx + \lambda_2\int_\Omega |I(x)-c_2|^2(1-H(\phi(x)))dx, 
\end{multline}
\noindent
where the term following $\mu$ represent the length of the contour, the term following $\nu$ the area inside the contour and $\delta$ the Dirac mass.
\begin{equation}
    (P): \arg \min_{c_1, c_2, \phi} F(c_1, c_2, \phi)
\end{equation}
\noindent
The new variable is $\phi$. The energy $F(c_1,c_2,\phi)$ is minimised wit respect to $\phi$ wit ha gradient descent evolution.

\paragraph{Chan-Vese Attention in U-Net architecture}

\begin{figure}[h]
\begin{center}
\begin{tcolorbox}[colback=white, colframe=black, arc=2mm, boxrule=0.5pt]
\includegraphics[width=11.6cm]{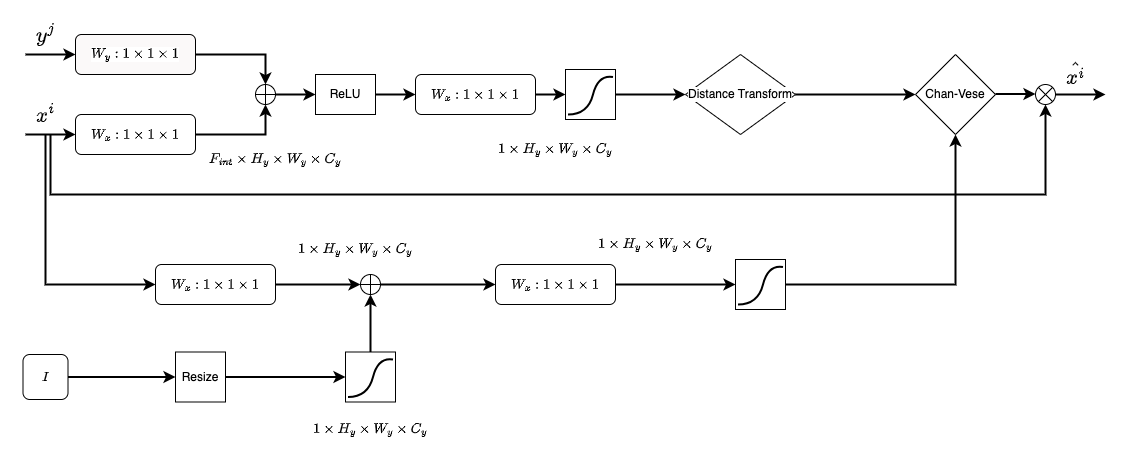}
\end{tcolorbox}
\end{center}
\caption{Scheme of the proposed attention method. The symbols $\bigoplus$ and $\bigotimes$  represent respectively the addition and multiplication of the tensors.}
\label{cv_attention}
\end{figure}

In this work, we incorporate our novel attention method, the $\textit{Chan-Vese Attention Gate}$, into the U-Net architecture. The attention method proposed by Oktay et al. \cite{1804.03999} allows highlighting areas that provide more information in the skip connections. Our method expands upon this approach by incorporating information from the Chan-Vese method, which performs intermediate segmentation for each block of the expansion phase. As illustrated in Figure \ref{cv_attention}, the representation constructed by the network from the skip connection is concatenated with the information from the previous layer before undergoing several transformations.
First, let us briefly review the classical attention gate proposed by Oktay et al. \cite{1804.03999}. Given a pair of input feature maps, the attention gate computes an attention coefficient that modulates the skip connections. The attention coefficient is obtained through a gating mechanism that involves two $1\times 1$ convolutions followed by an element-wise addition and a non-linear activation function (sigmoid). The attention gate effectively allows the network to learn which regions of the input image are more relevant to the segmentation task.
Our method builds upon the classical attention gate by incorporating the Chan-Vese segmentation technique. The first key step is the distance transform applied to the classical attention representation, which serves as the initial contour for the Chan-Vese method. This step is made possible thanks to the use of the distance transform method proposed by \cite{10.1007/978-3-030-71278-5_31}, which renders the transformation differentiable within a classical automatic differentiation framework. A secondary branch was implemented in the proposed framework to selectively emphasise the tumorous region of the input image. This branch carefully resizes the input image to conform to the dimensions of the current layer and integrates it with the residual feature map via addition. A $1\times 1$ convolution is then applied to the resulting map, which serves as the input for subsequent segmentation. In the second step, the modified Chan-Vese algorithm is used to iteratively segment the image a finite number of times. This segmented image is then used as a control signal to facilitate learning. The proposed approach thus provides a more refined and informative attention for the network.

Let $I \in\mathbb{R}^{N\times H\times W\times K_I}$ represent a batch of input images, where $N, H, W, K_I$ denote the batch size, height, width, and number of channels of the image, respectively. Let $X_i^f \in \mathbb{R}^{F_x \times H_l \times W_l \times K}$ denote the residual feature map $f$ at layer $i$ and $Y_j^f\in\mathbb{R}^{F_x\times H_l\times W_l\times K}$ the previous layer feature map at layer $j$ and feature map $f$. ${W_{1\times 1}}$ is a $1\times 1$ convolution. Following the additive attention formulation:

\begin{align}
& q_{att}^f = \Psi^T(\sigma_1(\boldsymbol{W_{1\times 1}^T}X_i^f + \boldsymbol{W_{1\times 1}^T}Y_j^f + b_f) + b_\psi \\
& \alpha_i^f = \sigma_2(q_{att}^f(X_i^f; Y_j^f;\Theta_{att}))
\end{align}
\noindent
We define $D_x$ as the distance transform that takes a tensor of shape $H_l \times W_l \times K_l$ as input. By applying the distance transform to the former attention gate, we obtain

\begin{equation}
\beta_i^l = D(\alpha_i^l) = -\lambda\log(\alpha_i^l \ast \exp(-\frac{d(\cdot,0)}{\lambda}))\
\end{equation}
\noindent
where $d(\cdot,\cdot)$ is the Euclidean distance and $\ast$ is the convolution product. This information is passed as an initialization contour.

On the other side, we perform a transformation of the input image, as if using some filter (e.g., CLAHE filter \cite{10.1016/S0734-189X(87)80186-X}), as follows:

\begin{equation}
\gamma_i^f = \sigma_2(\boldsymbol{W_{1\times 1}^T}(\boldsymbol{W_{1\times 1}^T}X_i^f + \sigma_2(I)) + b_W )\
\end{equation}
\noindent
Many areas of the images we wish to segment happen to have the same values as the averages of regions $c_1$ and $c_2$ as the region we ultimately want to segment. This sometimes leads the algorithm to add undesirable areas to the segmentation even though the $\mu$ and $\nu$ hyperparameters have been carefully chosen. The $\gamma_i^f$ transformation solves this problem by reducing the intensity in areas far from the tumour.

Finally, the resulting mask segmentation and attention coefficient $\zeta_i^f$ is given by solving the Chan-Vese problem:

\begin{equation}
\zeta_i^l = CV(\gamma_i^f, \beta_i^f, \mu, \nu)
\end{equation}
\noindent
where $\gamma_i^f$ is the image that supports the initial mask $\beta_i^f$. $\mu$ and $\nu$ are positive parameters.

During the backpropagation procedure, the gradients of the loss function concerning the attention gate's parameters (marked in bold) and intermediate outputs are computed using the chain rule, which can be efficiently executed in modern deep learning frameworks. The differentiable distance transform and the Chan-Vese module enable the backpropagation to update the parameters of the attention gate and the network's other layers. Consequently, the network can learn to focus on the most relevant regions for segmentation, improving its performance in medical image segmentation tasks.

\section{Experiments}
\label{sec:experiments}

\paragraph{Evaluation Datasets}
In this study, we used the TCGA\_LGG database, an openly available online repository \cite{tcga_lgg} containing magnetic resonance imaging (MRI) scans of brain tumour patients. The database consists of 110 patients from The Cancer Genome Atlas (TCGA) lower-grade glioma collection, with genomic cluster data and at least one fluid-attenuated inversion recovery (FLAIR) sequence available. Table \ref{tab:results} provides a summary of the experimental results.The spatial resolution of the images contained in the TCGA LGG dataset is $1mm$ isotropic.


\begin{table}[]
    \centering
    \caption{Segmentation results (IOU) on the TGCA\_LGG brain MRI database. Significant results are highlighted in bold font.}\label{tab:results}
    \begin{tabular}{|p{5em}|P{10em}|P{10em}|P{10em}|}
    \hline
         Method & U-Net & Attention U-Net & Chan-Vese U-Net  \\
         \hline
         Dice &  $\mathbf{0.832 \pm 0.091}$ & $0.830 \pm 0.023$ & $0.824 \pm 0.019$ \\
         IOU & $0.829 \pm 0.075$ & $0.833 \pm 0.023$ & $\mathbf{0.848 \pm 0.021}$ \\
         Hausdorff & $2.390 mm \pm 0.985$ & $2.416 mm\pm 0.775$  & $\mathbf{2.329 mm\pm 0.672}$ \\
         FPR & $0.010 \pm 0.003$& $\mathbf{0.009 \pm 0.002}$ &$0.012 \pm 0.004$ \\
         FNR & $0.013 \pm 0.004$ & $0.015 \pm 0.005$ & $\mathbf{0.013 \pm 0.003}$ \\
         \hline
    \end{tabular}
    
    \label{tab:results}
\end{table}

\paragraph{Implementation Details}
We used a large batch of 32 for gradient update and the model parameters are optimised using an adamW optimiser \cite{adamw} with learning rate $5\times 10^{-4}$ and batch normalisation. We applied standard data augmentation (resize, horizontal flip, vertical flip, random rotate, transpose, shift and scale, normalise). The Chan-Vese parameters $\mu$ and $\nu$ are set respectively to $0.1$ and $1.0$. The loss is computed using the addition of Dice loss and Binary Cross Entropy. The added attention layer slows down the training by an average of 1 sec out of 6 sec per batch. The code is written in Jax using Haiku framework and will soon be available.

\paragraph{Segmentation Results} In this study, we conducted a comparative evaluation of our proposed model with the classical U-Net and the original Attention U-Net. Table \ref{tab:results} provides a summary of the experimental results. Dice, IOU (Intersection Over Union), Hausdorff, FPR (False Positive Rate), and FNR (False Negative Rate) values are reported, along with their respective standard deviations (denoted as $sd_i$). The image resolution should be considered when interpreting the Hausdorff distance values. Our proposed model demonstrated superior IOU scores and improved false negative performance. This can be attributed to the model's ability to focus on a smaller area of interest and the integration of the Chan-Vese method, which enables more effective capture of relevant information and reduces the risk of information loss.

\paragraph{Chan-Vese Attention Masks analysis}

\begin{figure}[ht]
\begin{center}
\includegraphics[width=12.2cm]{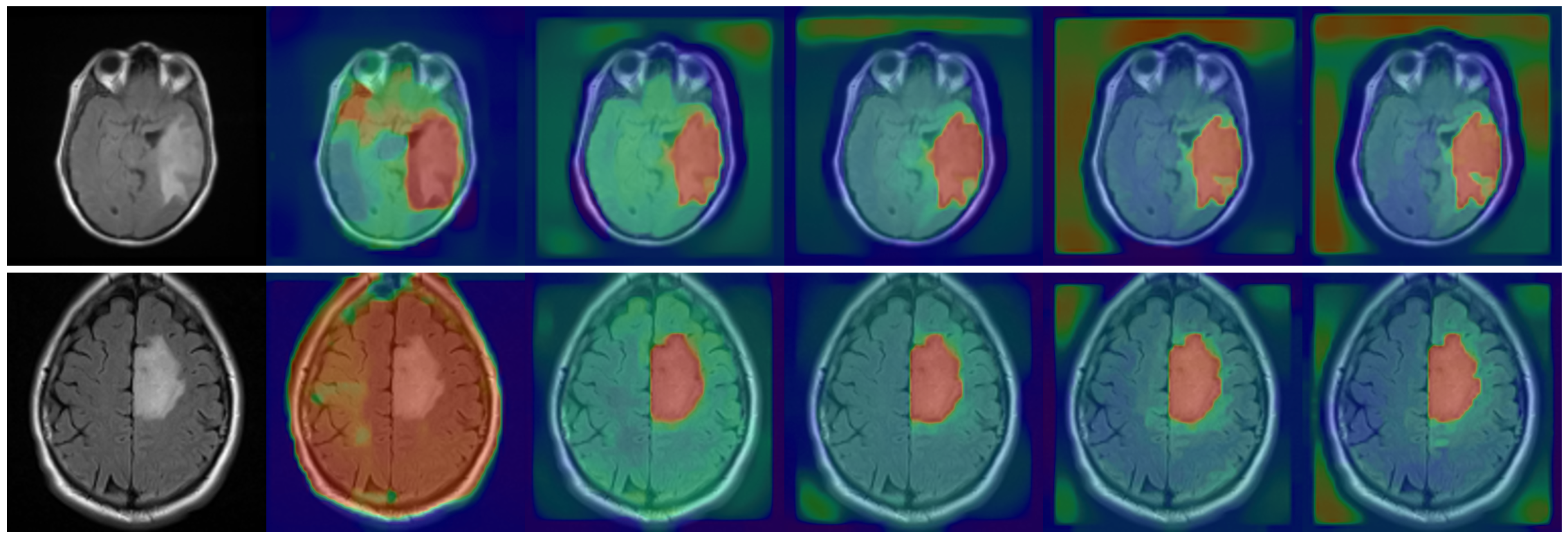}
\end{center}
\caption{Output of the Chan-Vese Attention Layer at different stage of learning (epochs:1, 50, 100, 200, 300) from left to right.}
\label{attention}
\end{figure}

We can observe from the results of the attention layer (see Figure \ref{attention}) that with the use of Chan-Vese the attention mask quickly converges to an apparently tumour-like segmentation. It takes advantage of the minimization of the Chan-Vese energy from the initialization of the mask thanks to the part inspired by the attention method of \cite{1804.03999} but also of the use of the initial image to be segmented. Gradually the contours of the tumour become more precise and the active intensity on the tumour is the confidence on the energy to be minimised. The 0 level set was used to enable the neural network to selectively prioritise the tumour area during segmentation. The upper level set was subsequently employed to refine the tumour segmentation. 

\paragraph{Comparison with Attention U-Net}

\begin{figure}[ht]
\begin{center}
\includegraphics[width=12.2cm]{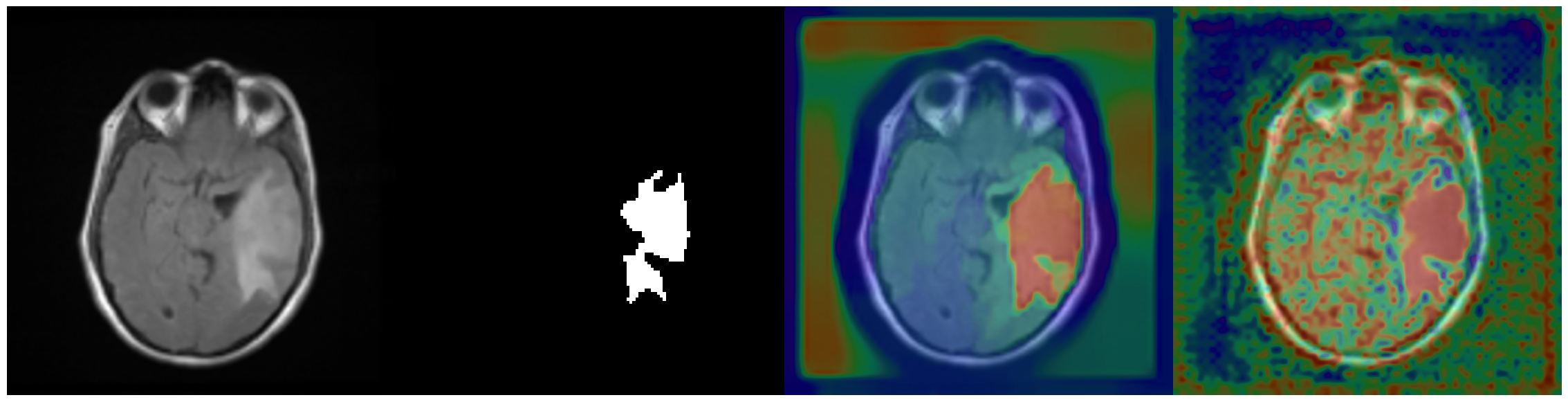}
\end{center}
\caption{Comparison of the Attention Mask between a Chan-Vese Attention and the Original Attention. (From left to right: the input MRI, the tumour to be segmented, Chan-Vese Attention mask, original Attention mask).}
\label{comparison}
\end{figure}

Figure \ref{comparison} shows the attention output of the Chan-Vese Attention Module and the classical Attention Module. Both methods allow the neural network to focus on the tumour area. It should be noted that the method proposed by \cite{1804.03999} obtains a finer mask on certain details of the tumour but does not manage in the framework of our study to rank the confidence of the presence of the tumour. In many places outside the tumour area we observe artefacts that do not correspond to the object of interest in the image. In contrast to these observations the method supported by Chan-Vese focuses only on the tumour area inside the skull.

\section{Conclusion}
\label{sec:conclusion}

In this paper, we have presented a novel segmentation approach that effectively combines classical energy minimization techniques with Deep Learning. Our proposed model, which integrates the Chan-Vese algorithm into the attention mechanism of a U-Net, demonstrates the value of incorporating non-deep learning sources of shape and structure information, particularly when dealing with sensitive medical data.

Our experimental analysis reveals that the proposed model achieves close results, with some improvements compared to both the classical U-Net and Attention U-Net in key performance metrics such as IOU scores and false negative rates. The Chan-Vese Attention Module successfully narrows down the model's focus to the tumour area, contributing to enhanced segmentation accuracy and precision. This work highlights the potential of combining diverse segmentation methods, paving the way for future research in blending various sources of shape and structure information with Deep Learning models. Ultimately, our approach aims to contribute to the development of more effective and versatile segmentation solutions.

\subsubsection{Acknowledgements} 
This work is in part supported by the French government under management of Agence Nationale de la Recherche as part of the "Investissements d’avenir" program, reference ANR-19-P3IA-0001 (PRAIRIE 3IA Institute).

\bibliographystyle{splncs04}
\bibliography{bibliography}

\end{document}